# Thermodynamic limit of the density matrix renormalization for the spin-1 Heisenberg chain

Stellan Östlund and Stefan Rommer

*Institute of Theoretical Physics*

*Chalmers University of Technology*

*S-41296 Göteborg, Sweden*

(March 17, 1995)

## Abstract

The density matrix renormalization group ("DMRG") discovered by White has shown to be a powerful method to understand the properties of many one dimensional quantum systems. In the case where renormalization eventually converges to a fixed point we show that quantum states in the thermodynamic limit with periodic boundary conditions can be simply represented by a special type of product ground state with a natural description of Bloch states of elementary excitations that are spin-1 solitons. We then observe that these states can be rederived through a simple variational ansatz making no reference to a renormalization construction. The method is tested on the spin-1 Heisenberg model.

In the past two years the density matrix renormalization group ("DMRG") method has been extensively tested on 1-d spin systems and spectacular numerical accuracy of both ground state energies and elementary excitations have been obtained with modest numerical effort. [1–4] In contrast to many of these efforts we explore the nature of the DMRG construction using a relatively few number of basis states to keep the numerical calculations simple. We have then been able to describe the thermodynamic limit of the ground state and single particle excitations in a way that generalizes very simply to arbitrary numbers of states. Our work shows that the fixed point limit of the DMRG leads to an ansatz form for the ground state and elementary excited states which can be explored variationally and is fundamentally independent of a renormalization scheme.

For definiteness, we focus on the spin-1 antiferromagnetic Heisenberg chain with bi-quadratic interactions

$$H = \sum_{n=0}^{N-1} S_n \cdot S_{n+1} - \beta(S_n \cdot S_{n+1})^2.$$

In the DMRG scheme, we work recursively with blocks representing say $n$ sites numbered from the left on which a set of states $\{|\alpha\rangle\}$ are defined. In principle there are $3^n$ quantum states in the block but we keep only a set of $n_s$ "important" states in our basis which we expect accurately describe most operators in the ground state.

In the DMRG recursion the next spin to the right of the block is absorbed into a new block which now has the $3n_s$ states of the product representation $\{|\alpha\rangle_{n-1} \otimes |s_n\rangle\}$. The DMRG now provides a method to construct the projection operator $A_n{}^{\alpha,(\beta,s_n)}$ which projects these states to a set of new basis states now representing the important $n_s'$ states of the larger block. This is written

$$|\alpha\rangle_n = \sum_{\beta,s_n} A_n{}^{\alpha,(\beta,s_n)} |s_n\rangle \otimes |\beta\rangle_{n-1} \tag{1}$$

where $s_n$ is the z component of spin at site $n$. Steven White's DMRG algorithm is a particularly effective way to compute a good projection operator.

Two crucial points follow. First we perform a trivial change in notation: $A_n^{\alpha,\beta}[s_n] \equiv$

$A_n{}^{\alpha,(\beta,s_n)}$. Second, we observe that if a recursion fixed point exists for $A$ i.e. $A_n[s] \to A[s]$ as $n \to \infty$, we find $|\alpha\rangle_n =$

$$\sum_{s_n,s_{n-1}\ldots} (A[s_n]A[s_{n-1}]\ldots A[s_1])^{\alpha,\beta}|s_n s_{n-1}\ldots s_1\rangle \otimes |\beta\rangle_0$$

where $|\beta\rangle_0$ represents some state far away from $n$. The form of a wave function homogeneous in the bulk of the chain is now clear: For every $n_s \times n_s$ matrix $Q$, we define

$$|Q) = \sum_{\{s_j\}} tr(\, QA[s_n]A[s_{n-1}]\ldots A[s_1]\, )|s_n s_{n-1}\ldots s_1\rangle. \qquad (2)$$

Thus $|Q)$ represents a state that is uniform in the bulk but a linear combination of boundary conditions defined by $|\alpha\rangle_n$ on the left and $|\beta\rangle_0$ on the right. [5,6] The special case of $Q = 1$ the identity matrix leads to a translationally independent state with periodic boundary conditions. Since this state turns out to be normalized we write $|1) = 1)$. For the "AKLT" $\beta = -1/3$ model our ground state ansatz is exact [7] as are the "matrix product" states. [8,9]

Several important properties of $A$ follow. First the projection should preserve orthonormal bases: $\delta_{\alpha,\alpha'} = \langle \alpha'|\alpha\rangle$. Using the recursion formula Eq. 1 and the orthogonality of the local spin states and previous block states, we find

$$\begin{aligned}\delta_{\alpha,\alpha'} &= \sum_{\beta\beta'ss'} A^{*\alpha',\beta'}[s']A^{\alpha,\beta}[s]\langle s'|s\rangle\langle \beta'|\beta\rangle \\ &= \sum_s (A[s]A^\dagger[s])^{\alpha,\alpha'}.\end{aligned} \qquad (3)$$

Hence in matrix form we find $\sum_s A[s]A^\dagger[s] = 1$.

We would also like our trial ground state with periodic boundary conditions to be an eigenstate of parity $\mathcal{P}$ where

$$\mathcal{P}|1) = \sum_{\{s_j\}} tr(\, A[s_n]\ldots A[s_1]\, )\mathcal{P}|s_n\ldots s_1\rangle \qquad (4)$$

$$= \sum_{\{s_j\}} tr(\, A[s_n]\ldots A[s_1]\, )|s_1\ldots s_n\rangle. \qquad (5)$$

A sufficient condition is that there exists an invertible matrix $Q_\mathcal{P}$ where

$$Q_\mathcal{P} A[s] = sign[\mathcal{P}](A[s])^T Q_\mathcal{P} \qquad (6)$$

where $A^T$ denotes transpose. Then $\mathcal{P}|1\rangle = (sign[\mathcal{P}])^n |1\rangle$. This assertion is shown by inserting $Q_\mathcal{P} Q_\mathcal{P}^{-1}$ into the trace in Eq. (5), then commuting $Q_\mathcal{P}$ through each matrix inside the trace and ultimately canceling the $Q_\mathcal{P}^{-1}$ again.

To find $Q_\mathcal{P}$ multiply each side of Eq. (6) by $A^\dagger[s]$ and sum over $s$. Using Eq. (3) we find that

$$Q_\mathcal{P}^{\alpha,\beta} = sign[\mathcal{P}] \sum_{s\tau\nu} (A^T[s])^{\alpha,\tau} Q_\mathcal{P}^{\tau,\nu} (A^T[s])^{\nu,\beta}$$
$$= sign[\mathcal{P}] \left( \sum_{s\tau\nu} (A^T[s])^{\alpha,\tau} A^{\beta,\nu}[s] \right) Q_\mathcal{P}^{\tau,\nu}$$
$$= sign[\mathcal{P}] \sum_{s\tau\nu} (A^T[s] \otimes A[s])^{(\alpha,\beta),(\tau,\nu)} Q_\mathcal{P}^{\tau,\nu}$$

Thus, $Q_\mathcal{P}$, if it exists, is the eigenvalue of the operator $\sum_s (A^T[s] \otimes A[s])$ with value $\pm 1$. In cases that we have looked at we have had no trouble finding $Q_\mathcal{P}$.

Our basis states should form a representation of total spin. Since adding an integral spin does not mix half odd or integral reps, the block can consist of reps that are all half-odd or all integer spin. Keeping approximately twelve states we have found that the representations of half integer spin gives far better numerical results, and in this case twelve states consisting of two spin 1/2 and two spin 3/2 in the total spin representations optimizes numerical accuracy.

In the diagram in Fig. 1 we have labeled on the left the "old" representations uniquely by the ordinal number $\gamma$ with the total spin $j$. New reps $\gamma'$ and total spin $j'$ are shown similarly on the right. Implicit in the labeling of the old states is total $j_z$ which is not shown. There are thus four irreducible representations of total spin with a total of twelve "old" states $|\gamma, j_z\rangle$.

When adding the new spin there are 36 "intermediate" states in the product representation falling into 10 different irreducible representations. We project from these reps back down to 4 irreducible "new" reps $\gamma'$. The nonzero projection terms $P^{\gamma',\gamma}$ are indicated by lines in Fig. 1. There are sixteen nonzero projection terms which are related by the requirement that our new states are orthonormal.

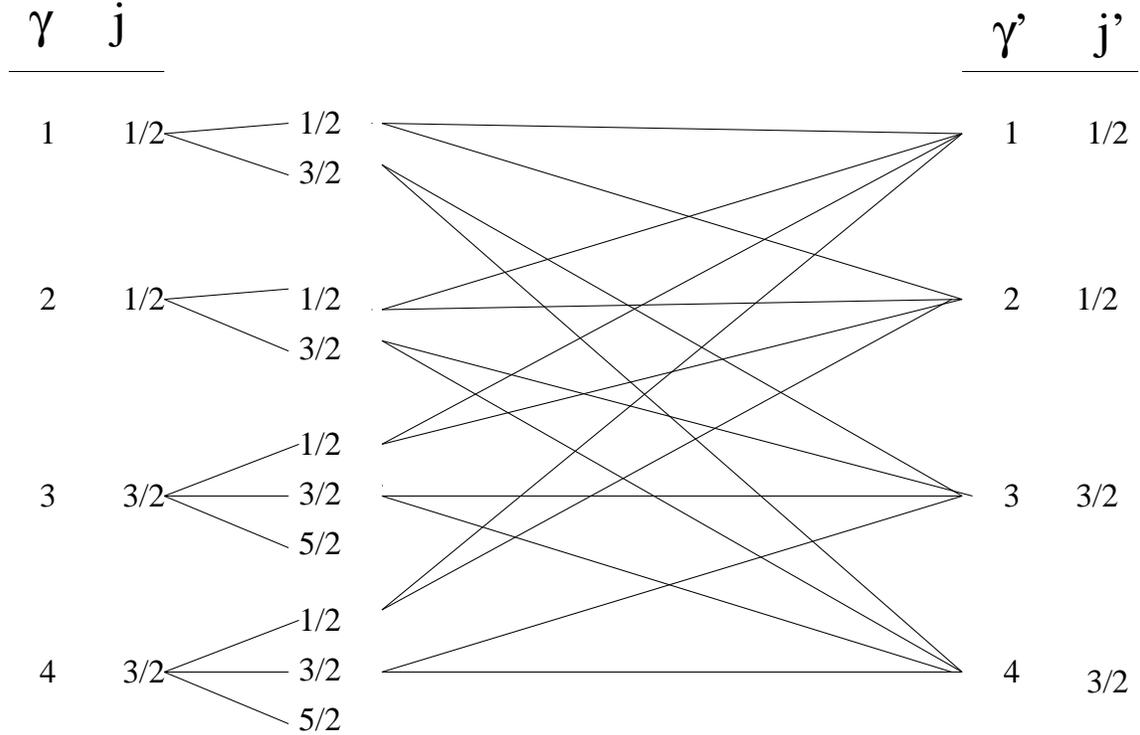

FIG. 1. The construction of the block states is shown when 12 states are kept in the basis. Old reps are on the left and new reps on the right. Each line represents a nonzero projection $P^{\gamma',\gamma}$ of basis reps.

Expressing this relation mathematically, we let $\gamma$ uniquely label a representation of total spin and $j(\gamma)$ denote the value of total spin. Each old state is thus labeled by $|\gamma, m\rangle$, where $m$ is the z-component of total spin. The new states are then given by

$$|\gamma', m'\rangle = \sum_{\gamma} P^{\gamma',\gamma} |\gamma, j(\gamma'), m'\rangle \qquad (7)$$

where $|\gamma, j(\gamma'), m'\rangle$ denotes the intermediate states formed by $|\gamma\rangle \otimes |s\rangle$ written using total-spin basis. Since we demand that the projection operators preserve total $j$ and $m$, these states can be explicitly constructed using the Clebsch-Gordan coefficients $\langle j_1, m_1|(j, m)(1, s)\rangle$ in the following form

$$|\gamma, j(\gamma'), m'\rangle = \sum_{m,s} \langle j(\gamma'), m'|(j(\gamma), m)(1, s)\rangle \left(|s\rangle \otimes |\gamma, m\rangle\right).$$

Inserting this into Eq. (7) we find that

$$|\gamma', m'\rangle = \sum_{s,(\gamma,m)} A^{(\gamma',m'),(\gamma,m)}[s](|s\rangle \otimes |\gamma, m\rangle)$$

where

$$A^{(\gamma',m'),(\gamma,m)}[s] = P^{\gamma',\gamma}\langle j(\gamma'), m'|(j(\gamma), m)\,(1, s)\rangle.$$

Thus, although the projection matrices $A$ contain $3 \times 12 \times 12$ numbers there are in fact only the few degrees of freedom available in $P^{\gamma',\gamma}$.

There are sixteen parameters in $P^{\gamma',\gamma}$. However demanding orthonormality of the basis states turn out to yield six constraints. Furthermore we fix the freedom of mixing arbitrarily the two spin 1/2 and likewise the spin 3/2, giving a total of only eight free parameters. [10] We can then use a variational principle for the energy to determine these parameters. At this point it is clear that the DMRG plays no essential role in the construction aside from providing a guide which reps to keep.

Expectation values of operators such as the Hamiltonian or correlations are given by sums of terms of the form

$$\langle 1|h|1\rangle = \sum_{\{s_j\}\{s_{j'}\}} tr(\,A^*[s_n']...A^*[s_1']\,) \times \\ tr(\,A[s_n]...A[s_1]\,)\langle s_n'...s_1'|h|s_n...s_1\rangle. \qquad (8)$$

To simplify this, we denote the identity for tensor products: $(A \otimes B)^{(\alpha,\beta),(\tau,\nu)} = A^{\alpha,\tau}B^{\beta,\nu}$ and the relation $tr(\,B\,)\,tr(\,A\,) = tr(\,B \otimes A\,)$. We define the mapping $\widehat{M}$ from $3 \times 3$ spin matrices $M$ to $n_s^2 \times n_s^2$ matrices by

$$\widehat{M} = \sum_{s's} M_{s',s} A^*[s'] \otimes A[s]$$

We let $S \equiv (S_x, S_y, S_z)$ define the spin-1 representation of total spin, and $\hat{1} \equiv \widehat{Id}$ the "hat" mapping of the $3 \times 3$ identity matrix. The operator $e^{i\pi S_z}$ is identified as the string operator. [11]

Using Eq. (8), we see that expectation values of spin, energy and spin correlation function are given respectively by

$$\langle 1|S|1\rangle = tr(\ \hat{1}^{n-1}\hat{S}\ )$$

$$\langle 1|S_j \cdot S_{j+1}|1\rangle = tr(\ \hat{1}^{n-2}\hat{S}\hat{S}\ )$$

$$\langle 1|S_j \cdot S_{j+m}|1\rangle = tr(\ \hat{1}^{n-m-1}\hat{S}\hat{1}^{m-1}\hat{S}\ ).$$

while the string order parameter correlation function is given by

$$\langle 1|S_0 e^{i\pi(S_1)_z}...e^{i\pi(S_{m-1})_z} S_m|1\rangle =$$
$$tr(\ \hat{1}^{n-m-1}\hat{S}(\widehat{e^{i\pi S_z}})^{m-1}\hat{S}\ ).$$

We note that these formulas are identical to those for one dimensional finite temperature classical statistical mechanical models, where the matrix $\hat{1}$ is identified as the transfer matrix and $S_j$ the ordinary spin operator.

The spectrum of correlation lengths, i.e. the collection of all possible exponential decay lengths $\xi$ of all correlation functions of the form $\langle \mathcal{O}_1(x)\mathcal{O}_2(y)\rangle \to ae^{-|x-y|/\xi}$ is given by the negative of the logarithms of eigenvalues of $\hat{1}$ with a similar relation of $\widehat{e^{i\pi S_z}}$ to the spectrum of string correlation lengths. We note that $\hat{1}$ is guaranteed to have an eigenvalue of 1 due to Eq. (3). Since the spin operator is orthogonal to this eigenvalue the next leading eigenvalue determines the decay of spin correlations. The eigenvalue of 1 in $\widehat{e^{i\pi S_z}}$ gives the long range order in string correlations. [12]

An ansatz for a Bloch state $|Q,k\rangle$ of momentum $k$ is given by $|Q,k\rangle_n =$

$$\sum_m e^{imk} tr(\ A[s_n]...A[s_{m+1}]QA[s_m]...A[s_1]\ )|s_n...s_1\rangle$$

where $Q$ is an arbitrary $n_s \times n_s$ matrix. Using the cyclicity of the trace, we find that

$$(Q',k|Q,k)_n = \sum_{m=0}^{n-1} tr(\ (Q\otimes 1)(\widehat{e^{ik}})^m(1\otimes Q')\hat{1}^{n-m}\ ) \qquad (9)$$

The Hamiltonian $H_{Q',Q}(k,n) = (Q',k|H|Q,k)_n$ can be calculated similarly. When $n$ is a power of 2 these sums can be computed recursively with an effort of $\log(n)$.

We define the discrete Laplace transform of a series $\{a_n\}_{n=0}^\infty$ by $F(\lambda) = \sum_{n=0}^\infty a_n e^{-n\lambda}$. The discrete Laplace transform of Eq. (9) and of the Hamiltonian as a function of $n$ is given by a convolution product. Given the numerical values of $\hat{1}$ and $\widehat{e^{i\pi S_z}}$ we can then obtain an analytic expression for the discrete Laplace transform and extract the leading behavior of $a_n$ as $n \to \infty$. Defining $H_{Q',Q}(k,n) \equiv (Q',k|H|Q,k)_n$ and $G_{Q',Q}(k,n) \equiv (Q',k|Q,k)_n$ it can simply be shown that when $k \neq 0$ for large $n$

$$H(k,n) = nH_1(k) + H_0(k) + \mathcal{O}(z)^n$$
$$G(k,n) = G_0(k) + \mathcal{O}(z)^n$$

where $|z| \approx .8 < 1$ represents the next-leading eigenvalue of the matrix $\hat{1}$. We can then solve for the ground state and excitation spectrum for a fixed value of $k$ in the thermodynamic limit if the following equation can be solved:

$$(nH_1(k) + H_0(k))|Q,k)_n = (nE_0 + \Delta_k)G_0(k)|Q,k)_n.$$

for all $n$ in the limit of large $n$. It can be shown that $H_1(k) \propto G_0(k)$, which allows us to identify the ground state energy $E_0$ per site through the proportionality $H_1(k) = E_0 G_0(k)$. The excitation spectrum $\Delta_k$ is then given by

$$H_0(k)|Q,k)_n = \Delta_k G_0(k)|Q,k)_n$$

Similar formulas can be obtained for $k = 0$. We note that the last formula is a Hamiltonian for the excitation spectrum which makes no explicit reference to a ground state.

These calculations have all been tested on the spin-1 Heisenberg chain, keeping the twelve states discussed before. The resultant 8-parameter family of trial ground states was explored to find the state of lowest energy for $\beta$ in the range 0 to 1. For $\beta = 0$ the variational ground state energy was found to be $E_0(\beta = 0) = -1.40138$, $E_0(\beta = .6) = -2.9184$ and $E_0(\beta = 1) = -3.98455$. The most accurate ground state calculations indicate that $E_0(\beta = 0) = -1.401484038971(4)$ [1,3]

The result for $\beta = 1$ is to be compared with the exact value of $-4$ obtained in Ref. [13]. The parity operator has been computed in all cases and it is verified that the ground state has parity $-1^N$ where $N$ is the number of sites.

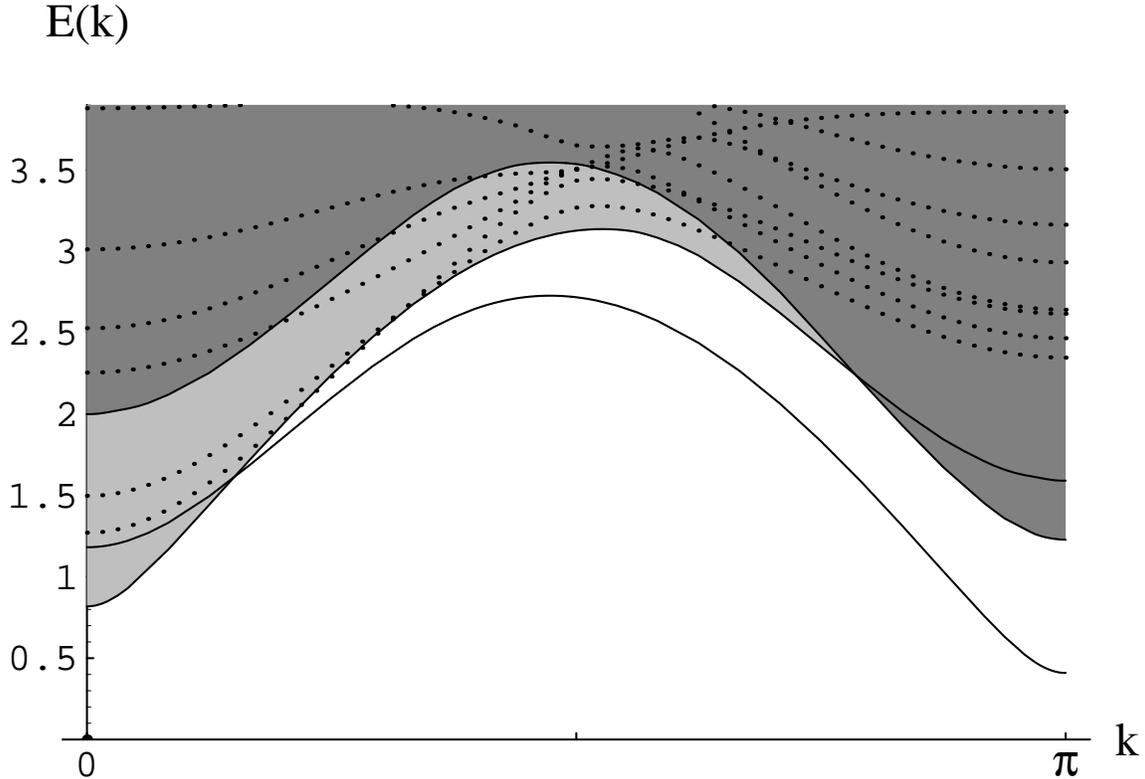

FIG. 2. The spectrum for $\beta = 0$ is shown. The lowest single particle triplet is shown as a solid line, with the lightly shaded region representing two-particle excitations and the dark region three particle excitations. Solid lines define the boundaries to the two and three particle continuum. Dotted lines indicate the spectrum of higher energy single-soliton excitations. The spin of these dotted excitations are, in order of increasing energy at $k = \pi$: 0,1,2,2,3,1,1,0.

An important issue is whether or not we have a good ansatz form for the excitations. For the pure Heisenberg chain $\beta = 0$, we find the single particle spectrum shown in Fig. 2. The low lying triplet branch defines the gap $\Delta_\pi = .4094$, which is very good compared to the most accurately known [1,3,14,15] result of 0.410502(1). Furthermore, we compute the spin wave velocity $v = 2.452$ to be compared to the calculations in Ref. [3] where $v = 2.49(1)$ was obtained. Clearly we reproduce the single particle triplet excitations with high accuracy considering the few number of states in our basis. Our calculation also yields a detailed spectrum of the lowest lying "single soliton" excitations shown by dotted lines in Fig. 2.

Our second lowest energy excitation at $k = \pi$ is a singlet shown by a dotted line in Fig. 2 with $\Delta_\pi(singlet) = 2.348$. As a function of $k$, the second lowest single-particle excitation is either a singlet or a spin-2 object, as has also been observed in exact finite size calculations. [16] For the string order parameter, we find $g(\infty) = -.3759$ whereas best estimates are [1] $g(\infty) = -.374325096(2)$. We find an asymptotic spin correlation length of $l = 3.963$ compared to best estimates of $l = 6.03$. The severe truncation of our basis to only twelve states has resulted in the asymptotic correlations being quite poor, although we have verified that *intermediate length* spin-spin correlations are consistent with more precise calculations. [17] Parity of each of the elementary excitations is verified by checking the relation Eq. (6) with $Q$ rather than the matrices $A$.

The boundary to two particle excitations at a given value of $k$ is shown in Fig. 2 computed explicitly by minimizing the sum of energies of excitations whose pseudomomentum sum to $k$, and similarly for the three particle excitations. These results are shown by the light shaded and dark shaded regions in Fig. 2. The picture fits well in with previously obtained results.

We have similarly computed spectra for various values of $\beta$. [18–20] Near $\beta = .6$ the excitation spectrum at $k = \pi$ crosses zero and becomes negative! Our interpretation of this is that our ground state ansatz is deficient, and this shows up as a condensation of elementary excitations. It is to be noted that Oitmaa et al. [16] also found that numerically the gap appeared to vanish rapidly near this value of $\beta$, although they too were unwilling to conclude that this persisted in the thermodynamic limit.

Our calculations are consistent with two possible scenarios of what happens near $\beta = .6$. A special value of $\beta$ could exist where the gap closes and signals a new phase. Or, the gap is in fact small and persists all the way to $\beta = 1$ but we do not see it due to our restricted ansatz for the ground state. In this case, we believe more accurate DMRG calculations will also have similar difficulties.

A significant issue appears to be that the DMRG fixed point seems to invariably lead to a matrix product ground state that, although it succeeds in reproducing ground state energies to high accuracy, cannot strictly give a power law decay of spin correlations. Thus, we find the energy very accurately at the Bethe ansatz point $\beta = 1$ without finding the expected powerlaw decay of correlations. The correlation length spectrum is given by the eigenvalues of the matrix $\hat{1}$, and it is hard to see how this can ever give algebraic correlations. However, intermediate correlations for intermediate lengths appear to be well represented in all cases.

# REFERENCES

[1] S.R. White and D.A. Huse, Phys. Rev. B **48**, 3844 (1993).

[2] S.R. White and R.M. Noack, Phys. Rev. Lett. **68**, 3487 (1993).

[3] E.S. Sørensen and I. Affleck, Phys. Rev. Lett. **71**, 1633 (1993).

[4] E.S. Sørensen and I. Affleck, UBCTP-93-026 (preprint).

[5] S. Quin, T.K Ng and Z.B Su, preprint (1995).

[6] K. Penc and H. Shiba, preprint cond-mat/9502085.

[7] I. Affleck, T. Kennedy, E.H. Lieb and H. Tasaki, Phys. Rev. Lett. **59**, 799 (1987).

[8] A. Klümper, A. Schadschneider and J. Zittartz, preprint.

[9] C. Lange, A. Klümper and J. Zittartz, cond-mat/9409107.

[10] In the even simpler case of saving only six states there are only two free parameters by similar arguments.

[11] M. den Nijs and K. Rommelse, Phys. Rev. B **40**, 4709 (1989).

[12] We have numerically observed that $\widehat{1}$ and $\widehat{e^{i\pi S_z}}$ have identical eigenvalue spectra but we have so far been unable to give a fundamental proof of this assertion.

[13] L.A. Takhtajan, Phys. Lett. **87 A**, 205 (1982); H.M. Babujian, Phys. Lett. **90 A**, 479 (1983); Nucl. Phys. B **215**, 317 (1983).

[14] G. Fáth and J. Sólyom, preprint.

[15] M. Takahashi, Phys. Rev. B **48**, 311 (1993).

[16] J. Oitmaa, J.B. Parkinson and J.C. Bonner, J. Phys. C **19**, L595 (1986).

[17] E.S. Sørensen and I. Affleck, UBCTP-93-022 (preprint).

[18] T. Xiang and G.A. Gehring, Phys. Rev. **48**, 303 (1993).

[19] R.J. Bursill, T. Xiang and G.A. Gehring, preprint cond-mat/9406076.

[20] R. Chitra *et al.*, preprint cond-mat/9412016.